**Comparative electrostatic force microscopy of tetra- and intra-molecular G4-DNA**

*Gideon I. Livshits*[1], *Jamal Ghabboun*[1†], *Natalia Borovok*[2], *Alexander B. Kotlyar*[2]*, *Danny Porath*[1]*

1 – Institute of Chemistry and The Harvey M. Krueger Center for Nanoscience and Nanotechnology, The Hebrew University of Jerusalem, Edmond J. Safra Campus, 91904 Jerusalem, Israel.
2 – Department of Biochemistry and Molecular Biology, George S. Wise Faculty of Life Sciences and The Center of Nanoscience and Nanotechnology, Tel Aviv University, Ramat Aviv 69978, Israel.

Gideon I. Livshits, Dr. Jamal Ghabboun, Prof. Danny Porath
Institute of Chemistry and The Harvey M. Krueger Center for Nanoscience and Nanotechnology, The Hebrew University of Jerusalem, Edmond J. Safra Campus, 91904 Jerusalem, Israel.
E-mail: danny.porath@mail.huji.ac.il
Dr. Natalia Borovok, Prof. Alexander B. Kotlyar
Department of Biochemistry and Molecular Biology, George S. Wise Faculty of Life Sciences and The Center of Nanoscience and Nanotechnology, Tel Aviv University, Ramat Aviv 69978, Israel.
E-mail: s2shak@post.tau.ac.il
[†]Current address: Department of Physics, Bethlehem University, Bethlehem, Palestinian Authority

The quest for conductive molecular nanowires for nanoelectronic devices has prompted the study of the electrical properties of DNA as a possible electrical conduit or template, primarily due to its molecular recognition and self-assembly properties.[1] Non-contact techniques, such as electrostatic force microscopy (EFM), can provide valuable information on the charge distribution, thus indicating on charge mobility, polarization and migration, within a single molecule by measuring its response to an external applied field with an oscillating probe above the molecule.[2-4] Here we report on comparative atomic force microscopy (AFM) and EFM measurements of two forms of guanine-based quadruplex DNA molecules, tetra- and intra-molecular G4-DNA. The tetra-molecular G4-DNA used in this work is made of four single-strands of guanine nucleotides that run parallel to each other[5]. Each strand is attached to a biotin molecule and four such strands are linked to an avidin tetramer. We label this type of tetra-molecular G4-DNA as BA-G4-DNA. Intra-molecular G4-DNA is obtained by self-folding of a single strand of guanines. Such folding leads to an anti-parallel configuration, in which two strands run in one direction and the other two strands run in the opposite direction.[6] When using the same number of tetrads for the construction of the tetra-molecular G4-DNA and the intra-molecular G4-DNA, the former are thicker and shorter than the latter molecules.[5] This suggests that the folding orientation of the strands, which form the backbone, affects the molecular structure, *i.e.* the tetrad unit and the tetrad-tetrad stacking. By comparing adjacent molecules of both types, co-adsorbed on the same mica surface, we circumvent the problem of phase calibration, showing that the EFM signal is twice as strong in the parallel configuration as compared with the anti-parallel G4-DNA, possibly because of greater charge mobility in tetra-molecular G4-DNA, thus making tetra-molecular G4-DNA a better candidate for conductivity measurements.

Theoretical[7, 8] and experimental[4, 9] studies showed that out of the four natural bases, guanine may form a $\pi$-stacking with the greatest chance of providing a conducting bridge between bases, due to its lowest oxidation potential.[9] Moreover, the robust quadruple helix, in which each tetrad (**Figure 1a**) is formed by eight hydrogen bonds rather than by two or three as in dsDNA, is more rigid than the duplex dsDNA helix and may withstand surface deformations in solid-state molecular devices. Such rigidity is particularly appealing for the realization of conducting molecular bridges. Previously, we reported the synthesis[6, 10] and EFM measurements[4] of intra-molecular G4-DNA (Figure 1b, left) which was stabilized by $K^+$ cations. This type of intra-molecular G4-DNA possessed distinct polarizability, in contrast to native dsDNA, which gave no discernible signal.[4] That study was followed by the synthesis and EFM measurements of tetra-molecular G4-DNA[5] (Figure 1b, right), which is stable in the absence of metal cations, and was shown to possess a clear signal of polarizability as well.

For intra-molecular and tetra-molecular G4-DNA that are composed of the same number of tetrads, AFM imaging[5] showed that tetra-molecular G4-DNA, adsorbed on a mica surface, resembles a tadpole with a large avidin “head” and a straight elongated “tail” (Figure 1c), while intra-molecular G4-DNA appears longer and thinner and is more flexible on the surface, forming curved lines. In this study, we synthesized two pairs of corresponding lengths of tetra-molecular G4-DNA and intra-molecular G4-DNA (see Figure 1c and **Figure S1**). Tetra-molecular G4-DNA was composed from either four 850 base-pairs (bp) (“short”) or 1400 bp (“long”) 5′biotin–poly(dG)-poly(dC) molecules attached to a single avidin as starting material (see Experimental Section). These molecules were 225 ± 20 nm and 280 ± 30 nm long, respectively, and with an average height of 2.2 ± 0.3 nm and 1.8 ± 0.2 nm, respectively (see Figure S1 for images and histograms). Intra-molecular G4-DNA, composed of a single strand of guanine, either 3400 bp (short) or 5500 kbp (long) long, formed a folded structure of nominally 850 and 1375 tetrads, respectively. The average length of these molecules is 250 nm ± 20 nm and 420 nm ± 30 nm long, respectively, with an average height of 1.1 ± 0.3 nm and 1.0 ± 0.1 nm, respectively, as measured by AFM (Figure S1). Since soft biological matter such as DNA is likely to have different measured apparent height due to different interactions with the scanning tip or with the surface, it is instructive to evaluate the ratio of the measured apparent heights. The height ratio of intra-molecular G4-DNA to tetra-molecular BA-G4-DNA (obtained in the same scan on the same substrate) is 0.50 ± 0.07 and 0.55 ± 0.08 for short and long pairs, respectively (measured for 100 molecules of each type). See Figure 1 and Figure S1 for details.

EFM comparison between intra-molecular G4-DNA and tetra-molecular BA-G4-DNA, was performed by measuring the response of these two species to an external electric field applied by a metalized AFM tip. Both types of molecules were co-adsorbed on a mica substrate as described in the experimental section below. EFM was done in either retrace mode[4] or 3D mode.[2, 4] Retrace mode is a 2-pass mode, in which the probe was lifted (typically 30–40 nm) above the set-point height, beyond the range in which Van der Waals (VdW) interactions are dominant and where the observed phase shift signal is mainly related to long-range forces, e.g. electrostatic. A weak but visible signal was observed in the phase-shift, and at this lift height, a complete scan was made when the tip was biased at two different voltages, ±4 V, and a control scan for a bias of 0 V. Then the tip was lowered, typically in steps of 5 nm, and for each voltage, a topography scan was measured (at set point height) and the phase-retrace was measured at the pre-defined lift. Tip deflection, amplitude and phase shift were measured at each height as a function of the distance between the sample and the probe, in order to verify the actual distance of the tip and monitor its deflection (see **Figure S2** and **Figure S3** for more details). There was a distinct signal at both the positive and negative bias, which was not present at 0 V bias until the low lifts (typically, 0 – 10 nm) where VdW interactions began to appear in the phase scan. During retrace, the feedback was disabled, and the dynamic amplitude was corrected in order to maintain the oscillation amplitude and to reveal tip deflection upon bias application that could lead to a wrong read of the phase shift imaging (see Figure S2 for more details). Tip deflection was taken into account as the lower bound of the oscillation of the tip. In order to calculate the intensity of the signal, in most cases the phase image was rotated to obtain a vertical profile, and then averaged along the molecule's length.

In 3D mode a line was chosen perpendicular to the molecule's length, and by disabling the y-scan, the tip scanned repeatedly along this line and then was continuously lifted at a predefined speed from the set-point height to 40 nm above it. Thus, retrace mode provides data regarding the intensity of the signal along the entire molecule at a constant lift, while 3D mode provides data along the vertical axis at a constant line (checked for drift by consecutive topography imaging).

**Figure 2** provides an illustrative example of the many long tetra-molecular BA-G4-DNA molecules, which were scanned. Three individual tetra-molecular BA-G4-DNA molecules are shown in Figure 2a, captured in the same scan area. EFM in retrace mode was performed from 5 nm to 30 nm above the set-point. The intensity of the signal is slightly stronger under the negative bias, which is consistent with a slightly negative bias towards a negatively charged mica. A typical scan of the phase retrace is shown in Figures 2b and 2d at ±4 V and a control scan at 0 V in Figure 2c, at a lift of 25 nm above the set-point. Measurements in 3D mode were performed along the green line in Figure 2a. The results are shown in Figures 2e-2f for ±4 V, respectively. The corresponding cross-sections, shown in Figure 2g, were obtained by averaging each figure (Figure 2e and Figure 2f) over the scan time (y-direction). Fifty individual tetra-molecular BA-G4-DNA molecules were measured and analyzed in this fashion. While it is impossible to quantitatively compare results from different scans with different parameters, a typical behavior of the intensity as a function of the lift is shown in Figure 2h. With some variability, we observe that as the tip is lifted higher, the signal fades into the background and is usually indistinguishable from the noise at about 35 nm above the set-point height.

The same techniques and procedures were used to investigate samples of co-deposited pairs of tetra- and intra-molecular G4-DNA, both short and long. Taking advantage of the fact that both types of molecules are scanned in the same setup and on the same sample simultaneously, it is possible to compare the intensity of their respective signals for various lifts.

**Figure 3** shows typical results from EFM in retrace mode over a pair of co-deposited short tetra- and intra-molecular G4-DNA (Figure 3a). Tetra-molecular BA-G4-DNA possesses a stronger visible EFM signal, in both positive (Figures 3b) and negative (Figure 3d) bias at a lift of 20 nm above set-point. A control measurement was made at 0 V bias at the same lift (Figure 3c). The corresponding averaged cross-sections are shown in Figures 3e and 3f. Figure 3g shows the intensity of the EFM signal as a function of the lift at a bias of +4V. Note that tetra-molecular BA-G4-DNA consistently displays a stronger signal at all the measured heights.

Figure 3h displays a ratio of the intensities, $D = \Delta\phi(\text{G4-DNA}) / \Delta\phi(\text{BA-G4-DNA})$, at positive and negative bias as a function of the lift. Note that the ratio is always significantly less than unity, and within the margin of error, is 0.5 ± 0.2. We have analysed fifteen pairs (seven short and eight long) of tetra- and intra-molecular G4-DNA molecules in this fashion. More examples and detailed measurements are shown in the supporting information.

Assuming a linear response of the feedback system for small amplitudes, the phase shift is given by[11]:

$$\Delta \phi \approx -\frac{Q}{k}\sum_i \frac{\partial F_i}{\partial z}, \quad (1)$$

where $Q$ is the quality factor, $k$ is the effective spring constant of the lever and $\sum_i \partial F_i / \partial z$ represents the sum of the derivatives of all forces acting on the cantilever in the surface normal direction $z$. It is reasonable to assume that the ratio of the phase-shifts of the molecules for a given lift would cancel out the influence of the intrinsic setup parameters (excitation frequency, effective spring constant, etc.), leaving only a dimensionless quantity that is purely the outcome of the intensities of the respective interactions. The ratio of the EFM intensities of intra-molecular G4-DNA to tetra-molecular BA-G4-DNA is presented in **Figure 4**. Since these dimensionless ratios are assumed to be independent of the setup (see also **Figure S4**), we can compare values from different measurements. For both positive and negative bias, we find an average ratio of 0.5 ± 0.1, regardless of the length. This implies that, irrespective of the setup, tetra-molecular BA-G4-DNA is twice as polarizable as intra-molecular G4-DNA.

In the present study, there are no metallic ions in tetra-molecular BA-G4-DNA, and yet the signal from this molecule displays twice the intensity of intra-molecular G4-DNA stabilized with $Na^+$ cations. The height and tetrad-tetrad separation in this new form of quadruplex G4-DNA suggests it is superior to intra-molecular G4-DNA, endowed with a different quadruplex shape that produces the electrostatic signal. Di Felice *et al.*[8] have calculated the bandwidth of G4-DNA as a function of the tetrad-tetrad separation and conformation, showing significant increase when compressive strain is applied along the helical axis. Their findings indicate that structural changes may have an immense effect on the electronic properties of the molecules and on the bandwidth in particular. More recently, Lech and co-workers[12] have calculated electron-hole transfer rates as a function of the conformational changes induced by different orientations and tetrad topologies. They demonstrated a great variance in the electron-hole transfer rates within the G-tetrad stacks for different stacking geometries, and identified one structure that allowed for strong electronic coupling and enhanced molecular electric conductance.

Their calculations suggest, in accordance with our experimental results, that the orientation of the tetrads plays an important role within the π-stacks. Our results provide experimental evidence that supports the idea that directionality of the strands may affect conformation, and consequently may affect π-π stacking and charge mobility along the molecule. These results support the existence of delocalized states in tetra-molecular G4-DNA and suggest it as a better candidate for charge transport measurements.

**Experimental Section**

*DNA*: Tetra-molecular BA-G4-DNA and intra-molecular G4-DNA were prepared as described in our previous publications.[5, 10] Co-deposited samples were prepared by adsorbing BA-G4-DNA and then G4-DNA. Typically, a 20-40 µl of 1–2nM BA-G4-DNA in 50mM HEPES and 2mM $MgCl_2$ were incubated on freshly cleaved mica for 10 min, washed with distilled water and dried with nitrogen gas. Subsequently, G4-DNA was deposited on this sample with 2mM $MgCl_2$, and incubated for 5 min. The sample was thoroughly washed with distilled water, and dried with nitrogen gas.

*AFM tip*: Soft $Si_3N_4$ cantilevers (OMCL-RC800PSA, Olympus Optical Co., Ltd) of nominal force constant 0.3 $Nm^{-1}$, resonance frequency 67 – 69 kHz and tip radius 15–20 nm were used (the tip radius was measured with a scanning electron microscope). A uniform gold/palladium layer was sputter-coated (SC7640 Sputter Coater, Polaron Inc.), which produced conductive tips (apex radius ~30 - 40 nm), and reduced the frequency to 59-62 kHz.

*Electrical Characterization*: Samples, scanning parameters and results varied, but a visible EFM signal was typically observed already at 30 nm above set-point. The appearance and strength of the signal depend primarily on two factors: the sensitivity of the cantilever and the free amplitude. A Nanotec Electronica AFM (Nanotec S.L., Madrid) was used to carry out measurements for AFM and EFM as described in the main text. A scan of the topography was made in dynamic mode to locate molecules or pairs of molecules. Prior to activating retrace mode or 3D mode, the parameters (bias, the tip lift distance or lift range for 3D) were given. At the retrace scan the feedback was disabled, and a bias was applied to the tip while the tip was lifted to be in the electrostatic range. Image analysis was performed with WSxM software.[13]

**Supporting Information**
Supporting Information is available from the Wiley Online Library or from the author.

**Acknowledgements**
The authors thank Lev Tal-Or, Izhar Medalsy, Julio Gomez-Herrero, Luis Colchero, Rosa Di Felice, Leonid Gurevich, Dvir Rotem and Igor Brodsky for helpful discussions and technical assistance. This work was supported by European Commission through grants ‘DNA-based Nanowires’ (IST–2001-38951), ‘DNA-based Nanodevices’ (FP6-029192); the ESF COST MP0802; GIF Grant No.: I-892–190.10/2005; the Israel Science Foundation (grant no. 1145/10); the BSF grant 2006422; The Minerva Center for Bio-Hybrid complex systems, the French Ministry of External Affairs, the Israeli-Palestinian Science Organization and Friends of IPSO, USA (with funds donated by the Meyer Foundation), and the INNI program through “Hybrid Functional coatings and Printed Electronics” HUJI project.
Note: Figures 2 and 3 were revised after initial publication online

[1] N. C. Seeman, Nature 2003, 421, 427; R. G. Endres, D. L. Cox, R. R. P. Singh, Rev Mod Phys 2004, 76, 195; D. Porath, G. Cuniberti, R. Di Felice, Top Curr Chem 2004, 237, 183.
[2] C. Gomez-Navarro, A. Gil, M. Alvarez, P. J. De Pablo, F. Moreno-Herrero, I. Horcas, R. Fernandez-Sanchez, J. Colchero, J. Gomez-Herrero, A. M. Baro, Nanotechnology 2002, 13, 314.
[3] C. Gomez-Navarro, F. Moreno-Herrero, P. J. de Pablo, J. Colchero, J. Gomez-Herrero, A. M. Baro, P Natl Acad Sci USA 2002, 99, 8484.
[4] H. Cohen, T. Sapir, N. Borovok, T. Molotsky, R. Di Felice, A. B. Kotlyar, D. Porath, Nano Lett 2007, 7, 981.
[5] N. Borovok, N. Iram, D. Zikich, J. Ghabboun, G. I. Livshits, D. Porath, A. B. Kotlyar, Nucleic Acids Res 2008, 36, 5050.
[6] A. B. Kotlyar, N. Borovok, T. Molotsky, H. Cohen, E. Shapir, D. Porath, Adv Mater 2005, 17, 1901.
[7] A. Calzolari, R. Di Felice, E. Molinari, A. Garbesi, Appl Phys Lett 2002, 80, 3331; A. Calzolari, R. Di Felice, E. Molinari, A. Garbesi, J Phys Chem B 2004, 108, 2509; P. B. Woiczikowski, T. Kubar, R. Gutierrez, G. Cuniberti, M. Elstner, J Chem Phys 2010, 133, 035103; J. Jortner, M. Bixon, T. Langenbacher, M. E. Michel-Beyerle, P Natl Acad Sci USA 1998, 95, 12759.
[8] R. Di Felice, A. Calzolari, A. Garbesi, S. S. Alexandre, J. M. Soler, J Phys Chem B 2005, 109, 22301.
[9] S. Steenken, S. V. Jovanovic, J Am Chem Soc 1997, 119, 617.
[10] N. Borovok, T. Molotsky, J. Ghabboun, D. Porath, A. Kotlyar, Anal Biochem 2008, 374, 71.
[11] R. Garcia, R. Perez, Surf Sci Rep 2002, 47, 197.
[12] C. J. Lech, A. T. Phan, M. E. Michel-Beyerle, A. A. Voityuk, J Phys Chem B 2013, 117, 9851.
[13] I. Horcas, R. Fernandez, J. M. Gomez-Rodriguez, J. Colchero, J. Gomez-Herrero, A. M. Baro, Rev Sci Instrum 2007, 78, 013705.

**Figure 1.** G4-DNA. (a) A scheme of a single G4 tetrad. (b) Schemes of intra-molecular G4-DNA (left) and tetra-molecular BA-G4-DNA (right). (c) AFM image of a pair of short BA-G4-DNA (left) and G4-DNA (right) co-deposited on a mica substrate. Inset shows a cross-section along the green line, revealing that BA-G4-DNA is thicker than G4-DNA.

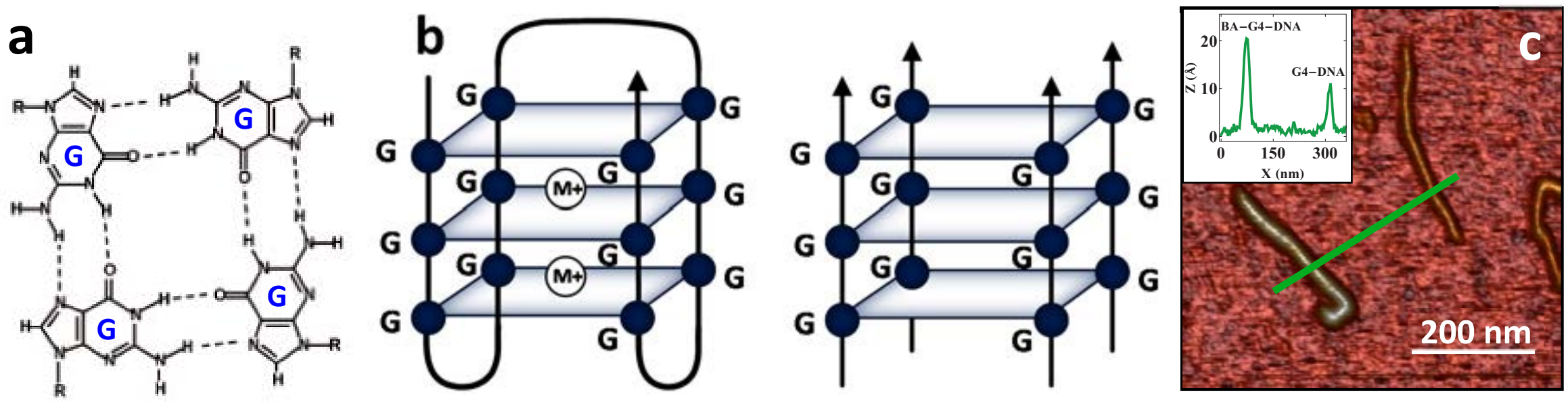

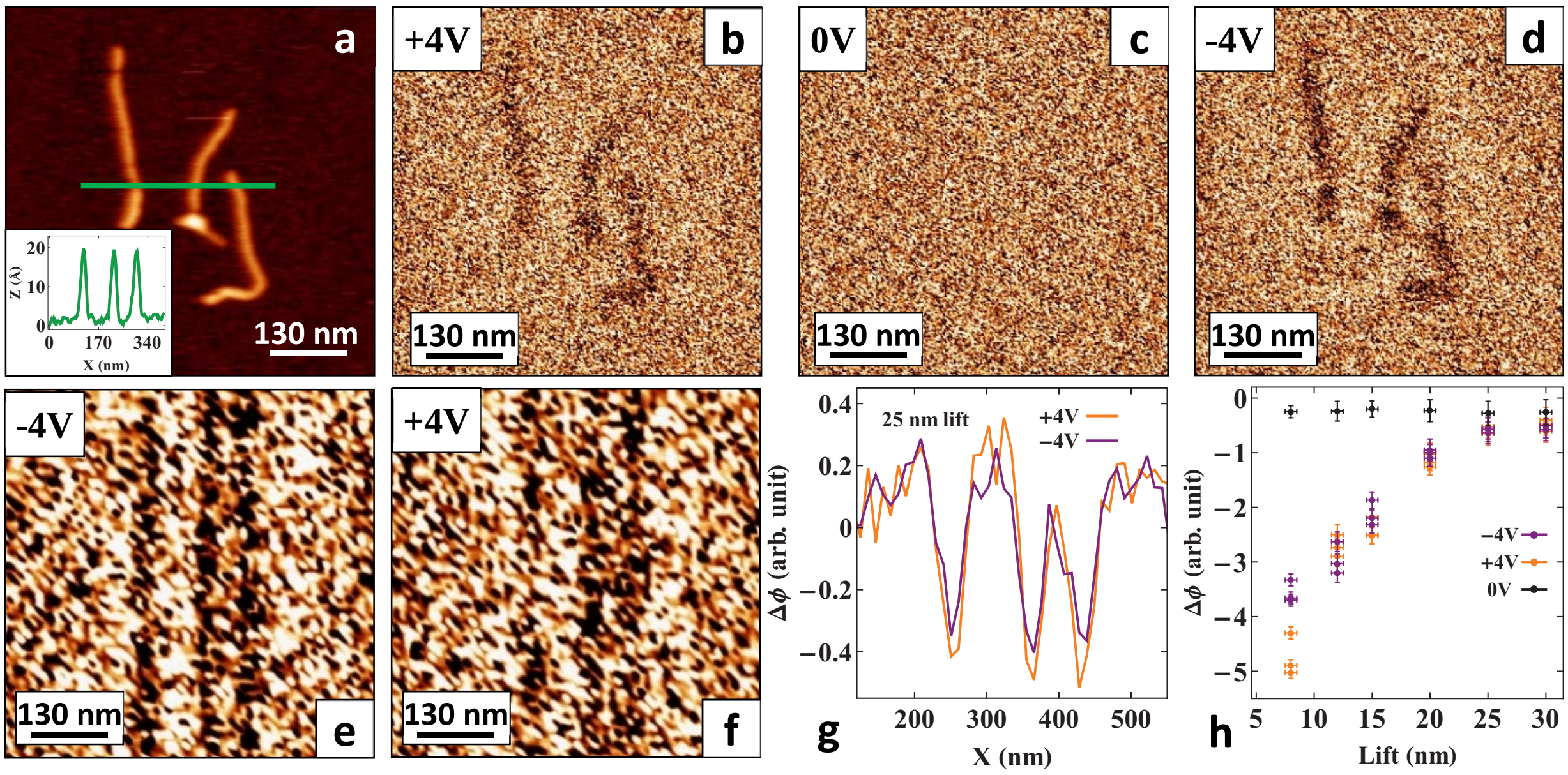

**Figure 2.** Typical comparative EFM measurement in retrace mode of a sample of long BA-G4-DNA. (a) AFM image of three BA-G4-DNA molecules collected while retracing; inset shows cross-section along the green line. (b), (c), (d) are phase-shift scans at 25 nm above set-point for +4, 0 and -4 V bias, respectively. (e) and (f) show phase-shift scans at a 25 nm lift obtained by disabling the y-scan along the green line in (a) for -4 V and +4 V, respectively. (g) The time-average of images (e) and (f). The phase shift is presented in arbitrary units. (h) The averaged intensity of each molecule is calculated, and is shown separately for each of the three molecules for positive (orange) and negative (purple) bias, along with the background noise (black). This demonstrates the variability in the response of similar molecules to the electric field, as well as the fact that as the signal fades, and variance is reduced as the tip-substrate distance (lift) is increased.

**Figure 3.** Comparative EFM imaging in retrace mode of a sample of co-deposited short BA-G4-DNA and G4-DNA. (a) AFM image measured while retracing. A pair of BA-G4-DNA (left) and G4-DNA (right) molecules is shown; the inset shows a cross section along the green line. (b), (c), (d) EFM scans at 20 nm above set-point for +4, 0 and -4 V bias, respectively. (e) and (f) show cross-sections of the averaged intensity for BA-G4-DNA (red oval) and G4-DNA (blue ovals) at 20 nm above set-point. Background noise (black) is obtained by averaging the areas between molecules. (g) EFM intensity as a function of the lift (G4-DNA in blue; BA-G4-DNA in red) at +4 V. (h) Dimensionless ratio of the EFM intensities of G4-DNA/BA-G4-DNA as a function of the lift at both positive (orange) and negative (purple) bias. In both **g** and **h**, errors in the horizontal axis are determined by system parameters and stability, while the error in the vertical axis represents background noise (g) at each lift, and corresponding error in (h).

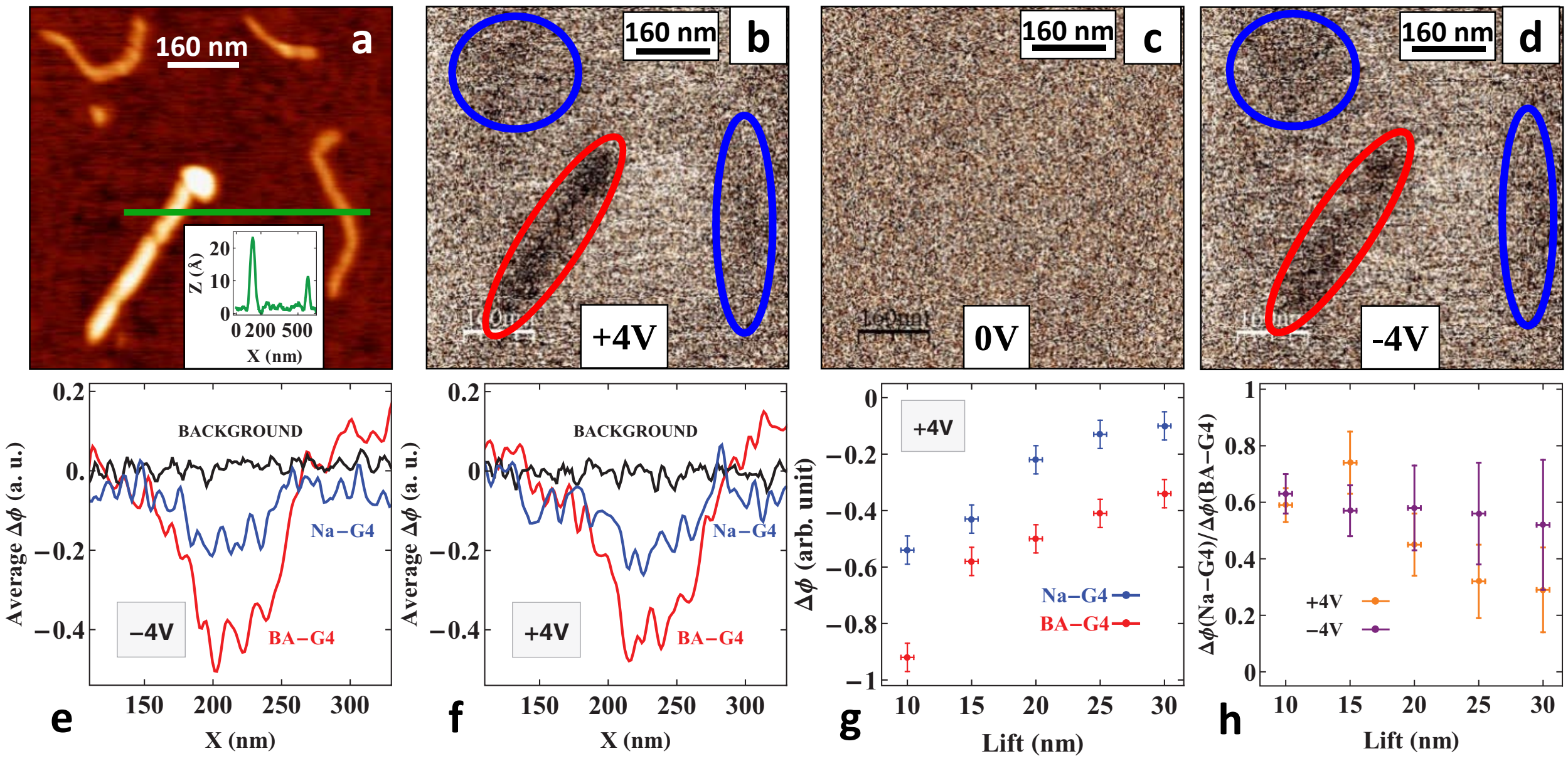

**Figure 4.** Ratio of intensities of the EFM signals of G4-DNA to BA-G4-DNA (similarly to Figure 3h), averaged over 15 pairs of molecules (both long and short) at +4 V (orange) and -4 V (purple). The ratio is dimensionless, and is system-independent. Dashed line (gray) shows the average value of 0.5, suggesting that, regardless of the length, on average BA-G4-DNA is twice as polarizable as G4-DNA. Errors in the horizontal axis are determined by system parameters and stability (see Figure S2 and Figure S3), while the error in the vertical axis represents the averaged error over different values of the ratio for corresponding pairs of molecules.

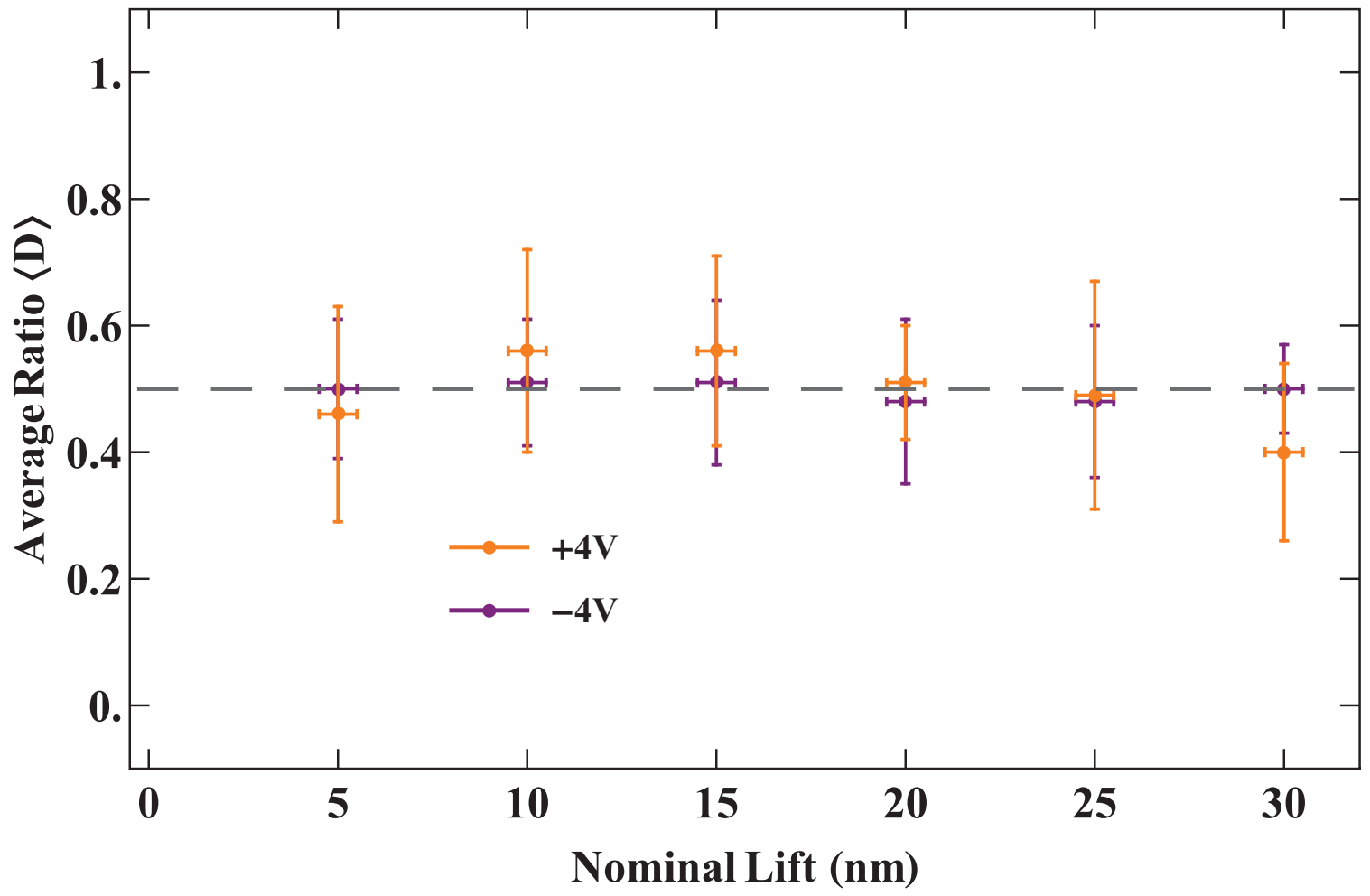

# Supporting Information

## Comparative electrostatic force microscopy of tetra- and intra-molecular G4-DNA

*Gideon I. Livshits, Jamal Ghabboun, Natalia Borovok, Alexander B. Kotlyar, Danny Porath**

This section contains additional controls and measurements cited in the main text. Figure S1 shows the relative height and length for the two pairs of synthesized lengths of tetra-molecular BA-G4-DNA and intra-molecular G4-DNA measured for 100 molecules in co-deposited samples. Figures S2, S3 and S4 constitute additional EFM and control measurements, primarily demonstrating controlled tip deflection with amplitude compensation during the application of bias voltage, consistent with the relative nominal lift in retrace mode.

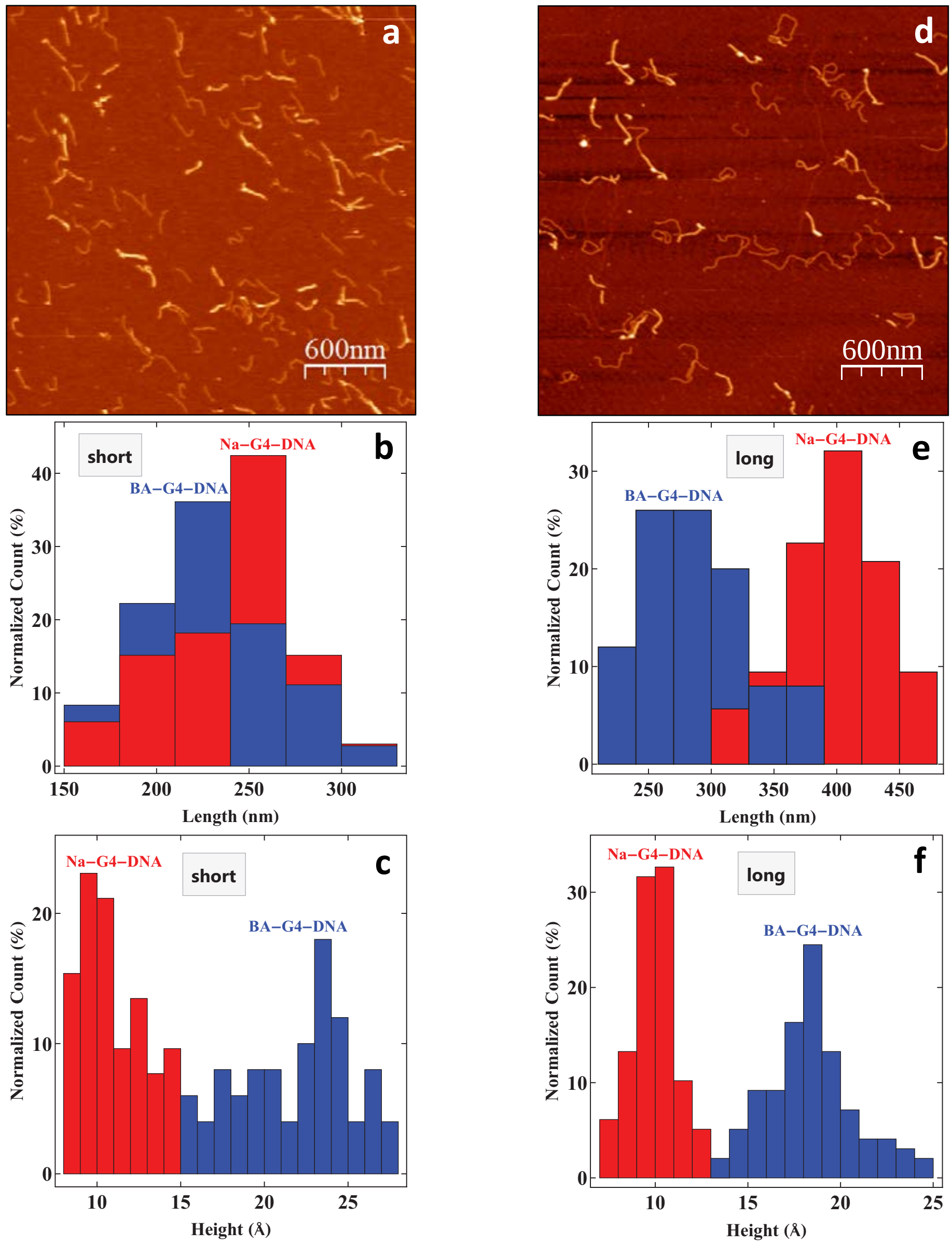

**Figure S1.** Comparative measurements of the length and height of BA-G4-DNA and Na-G4-DNA for the two pairs of synthesized lengths, 850 and 3400 bases ("short") and 1400 and 5500 bases ("long"), respectively. (a) AFM image of co-deposited short BA-G4-DNA and Na-G4-DNA molecules. (b) Normalized histogram of the corresponding lengths, showing that Na-G4-DNA is a bit longer, for the same synthesized length. Their respective lengths, 225 ± 20 nm and 250 ± 20 nm, are generally in accordance with the nominal distance between tetrads in short G-quadruplexes, 3.25 nm[1], although the Na-G4-DNA peak is a bit shorter. (c) Normalized histogram of the apparent height, showing that BA-G4-DNA is nearly twice as high as Na-G4-DNA, with a wide distribution of heights for both molecules. (d) AFM image of co-deposited long BA-G4-DNA and Na-G4-DNA molecules. (e) Normalized histogram of the corresponding lengths, showing that Na-G4-DNA length (420 nm ± 30 nm) is generally in accordance with the nominal distance between tetrads in short G4-quaruplexes, and is appreciably longer than BA-G4-DNA (280 ± 30 nm). (f) Normalized histogram of the apparent height, showing that long Na-G4-DNA is quite uniform in height, with a narrow distribution centered at 1.0 ± 0.1 nm. Long BA-G4-DNA is less uniform with a height of 1.8 ± 0.2 nm. The error in either height or length is determined by the width of the distribution at ~$e^{-1}$ of the peak value. Height values used in the histograms are an average of several height measurements along each molecule. Each histogram represents measurements of 100 molecules of each type. Where the histograms overlap, the bars are brought into the forefront.

**Figure S2.** Tip deflection and dynamic compensation of amplitude. (a), (b), (c) Plots of the tip deflection as a function of the tip-substrate separation, *z*, taken in retrace mode at a nominal lift of 20 nm above set-point. These three F-z curves correspond to an applied tip bias of 0 V, +4 V and -4 V, respectively. The forward (green) curve reveals tip oscillations that gradually diminish as the tip is brought closer to the substrate, eventually jumping into contact with the substrate (vertical dashed line in **a** and **b**, *z* = 57 nm). The application of bias in this setup creates an attractive electric force due to the polarization of charge on the substrate. The result is an uncontrollable deflection of the tip towards the substrate. This influences the actual distance of the tip from the substrate, as seen in **c**, where the amplitude was not adjusted unlike in a and b. Generally, to avoid this unwanted deflection – which may affect subsequent measurements – the amplitude is adjusted during retrace to maintain the same level of deflection and consistent height throughout the measurement. (d), (e), (f), F-z curves, obtained at a lift of 25 nm above set-point, in which the amplitude was dynamically adjusted. The forward (green) and backward (red) curves do not overlap, but the jump into contact is nearly the same for all applied tip biases (vertical dashed line, *z* = 5 nm), 0V, +4 V and -4V, respectively.

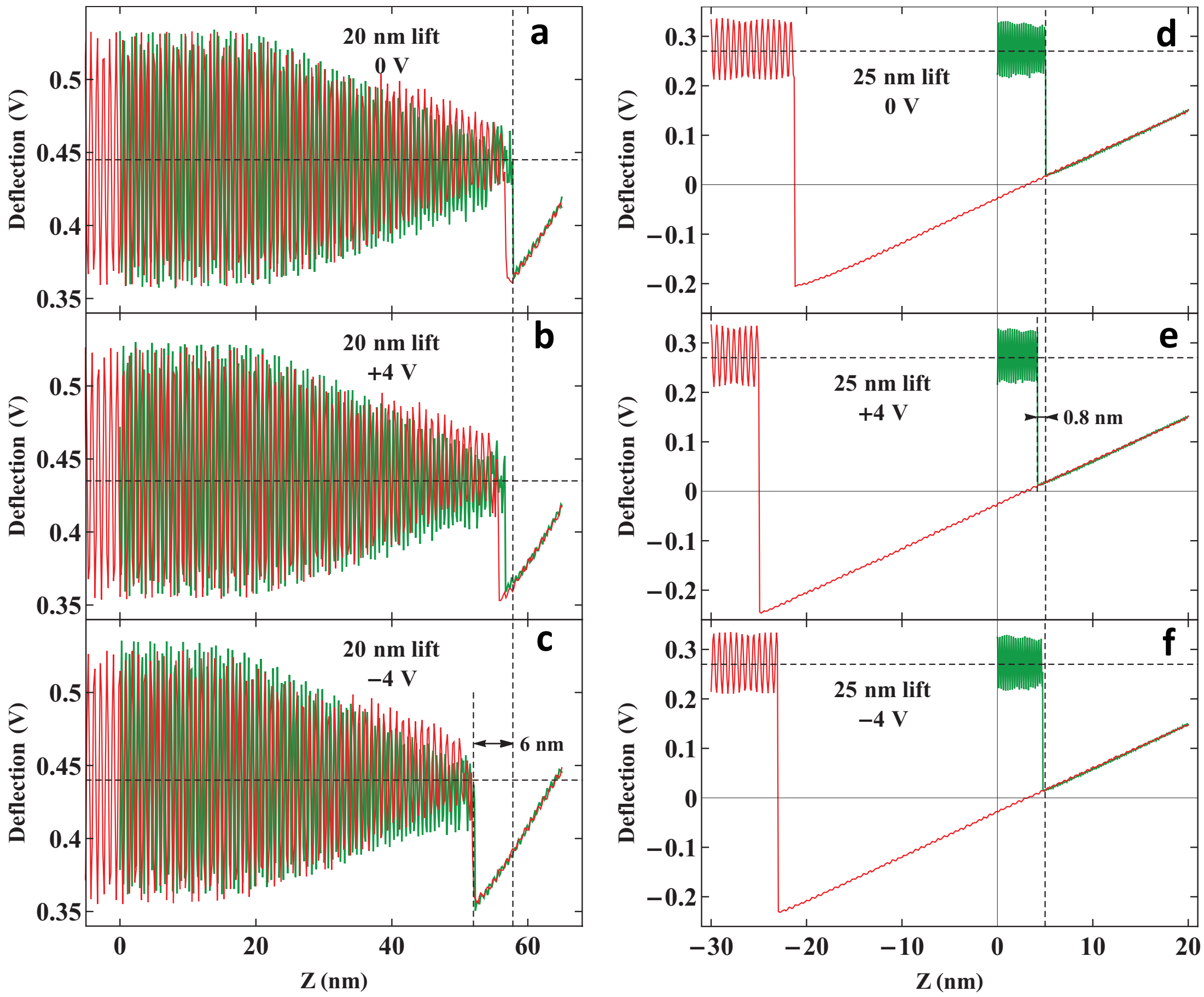

The series of measurements below is an example of a complete set of measurements in retrace mode. Each set is composed of AFM images taken at set-point height of a pair of co-deposited long tetra-molecular BA-G4-DNA and intra-molecular G4-DNA. At each lift, the phase retrace is measured for positive, negative and zero bias. Tip deflection, amplitude and phase as a function of the tip-sample distance, *z*, are measured at both set-point height and lift height. These plots are denoted by F-z, A-z and Ph-z, respectively. The results are shown in four separate panels, arranged into three columns according to the positive (left), zero (middle) and negative (right) bias applied to the tip during retrace.

**Figure S3. A.1 – A.15**, A set of comparative AFM and EFM measurements in retrace at a nominal lift of 20 nm above set-point. **A.1-A.3** are AFM images taken in dynamic mode at set-point height. BA-G4-DNA is clearly visible on the left, while G4-DNA is on the right. A segment of a second G4-DNA is visible on the top left as well. Cross sections of the molecules are shown in the inset. **A.4-A.6** are phase-retrace images, taken at the nominal height of 20 nm, at a bias of +4 V, 0 V and -4 V, respectively. A visible trace of both molecules is observed at both positive and negative bias. **A.7-A.9** Plots of forward F-z curves. These show only the forward direction at set-point height (green) and at the bias (blue). Since the amplitude is adjusted during retrace to compensate for uncontrollable tip deflection, the F-z curves may be used to deduce the actual tip-sample height as well as the deviation from the nominal height above set-point. A slight deviation of ±2 nm is observed. **A.10-A.12** Plots of the forward A-z curves. The amplitude during the lift in retrace is manually adjusted at the beginning of the scan to be the same as the amplitude at set-point height to compensate for the abrupt deflection of the tip. The effect of amplitude compensation is observed at both negative and positive bias, with only a slight deviation (~1-2 nm) in the amplitude at zero and -4 V. The A-z curves at zero bias and at set-point height are linear, while at ±4 V they display the well-known sigmoidal shape owing to the effect of the electrostatic force acting on the tip. **A.13-A.15** Plots of the forward Ph-z curves, showing a sudden change in the phase as the tip is brought into contact with the substrate. Panels **B** and **C** correspond to complete sets of measurements taken at the nominal lifts of 15 nm and 10 nm, respectively, while panel **D** shows retrace imaging taken at 5 nm above set-point where the VdW interaction is clearly visible (green circles) under zero bias. Panel **E** is the ratio of intensities for all lifts.

**Panel A: 20 nm lift**

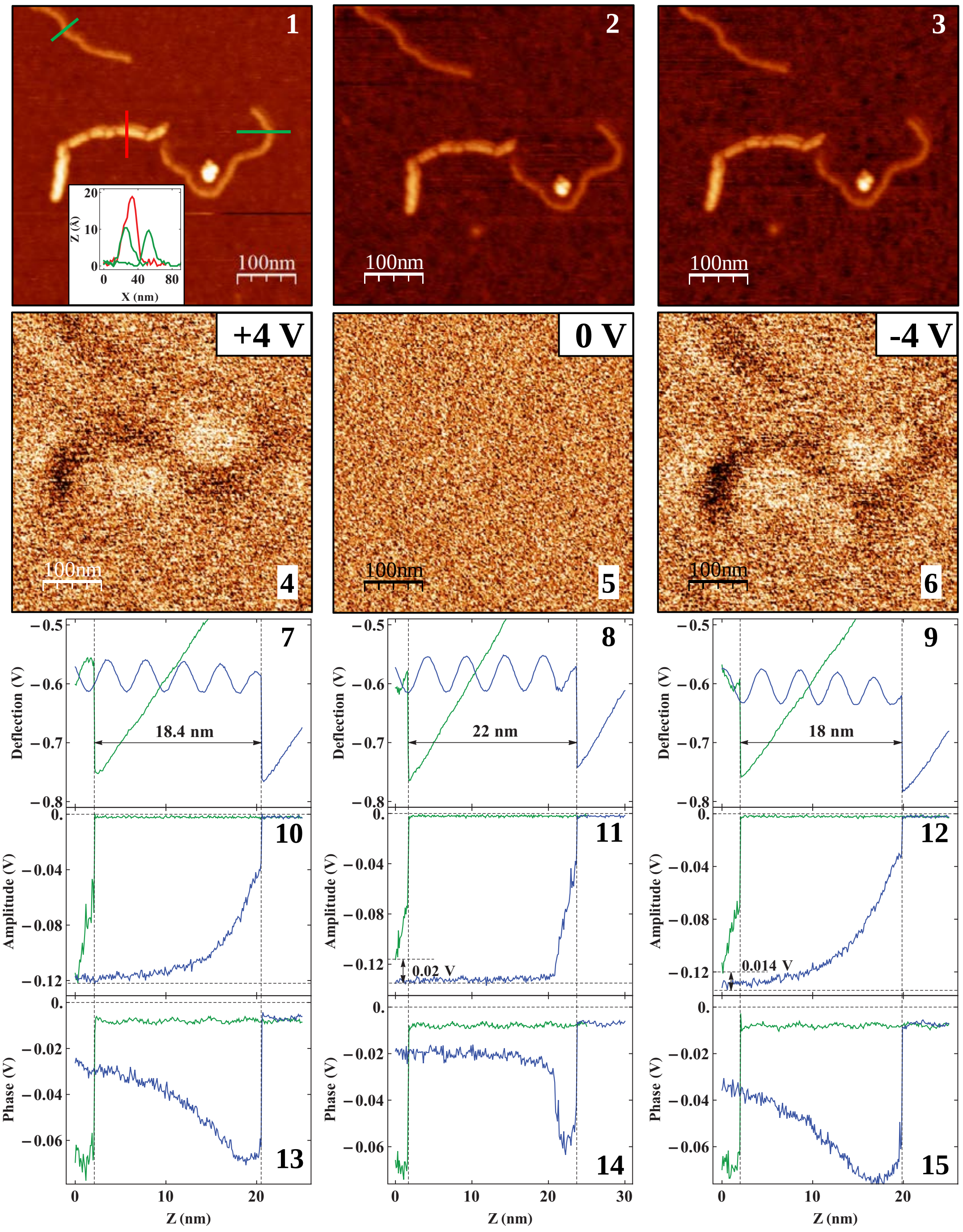

**Panel B: 15 nm lift**

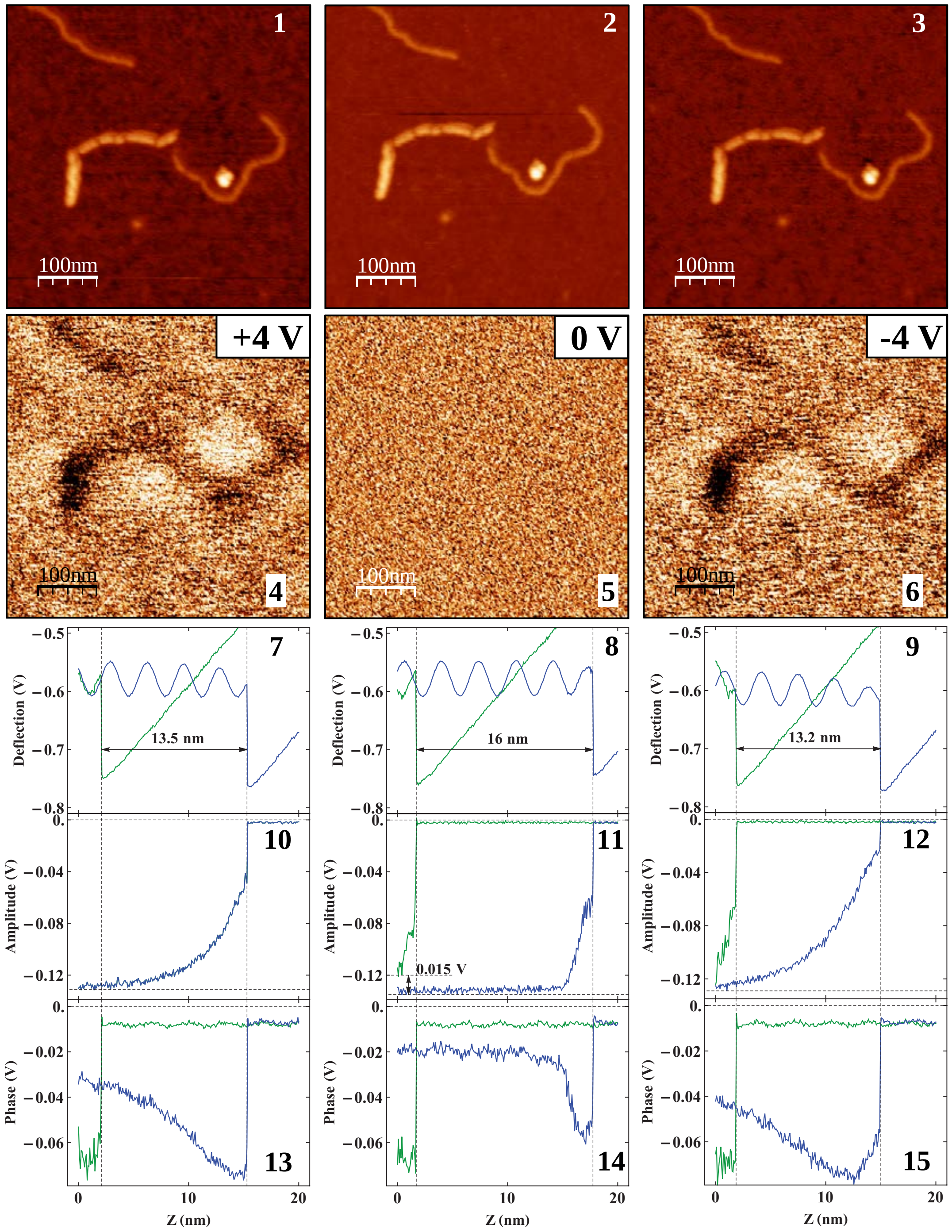

## **Panel C: 10 nm lift**

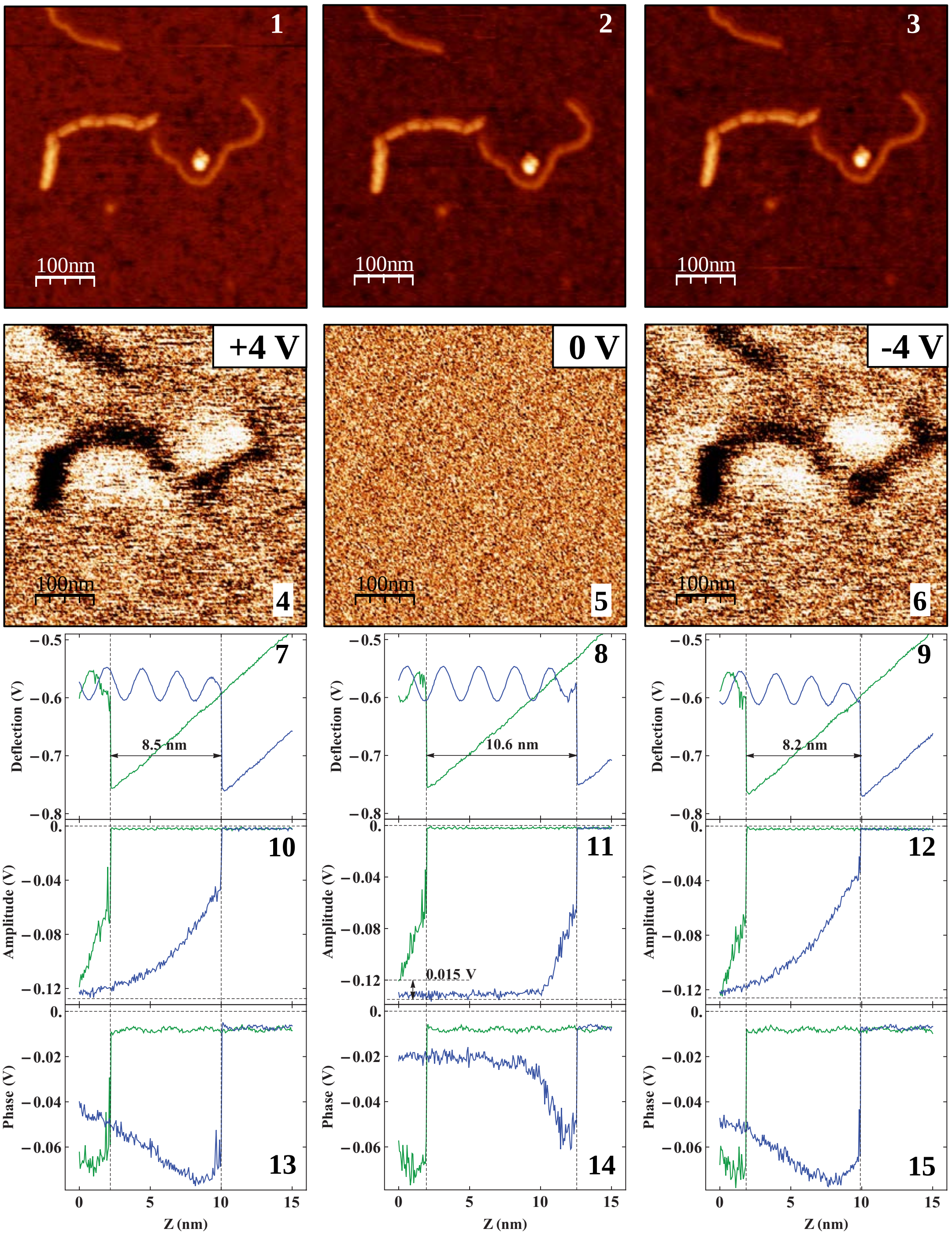

**Panel D: 5 nm lift**

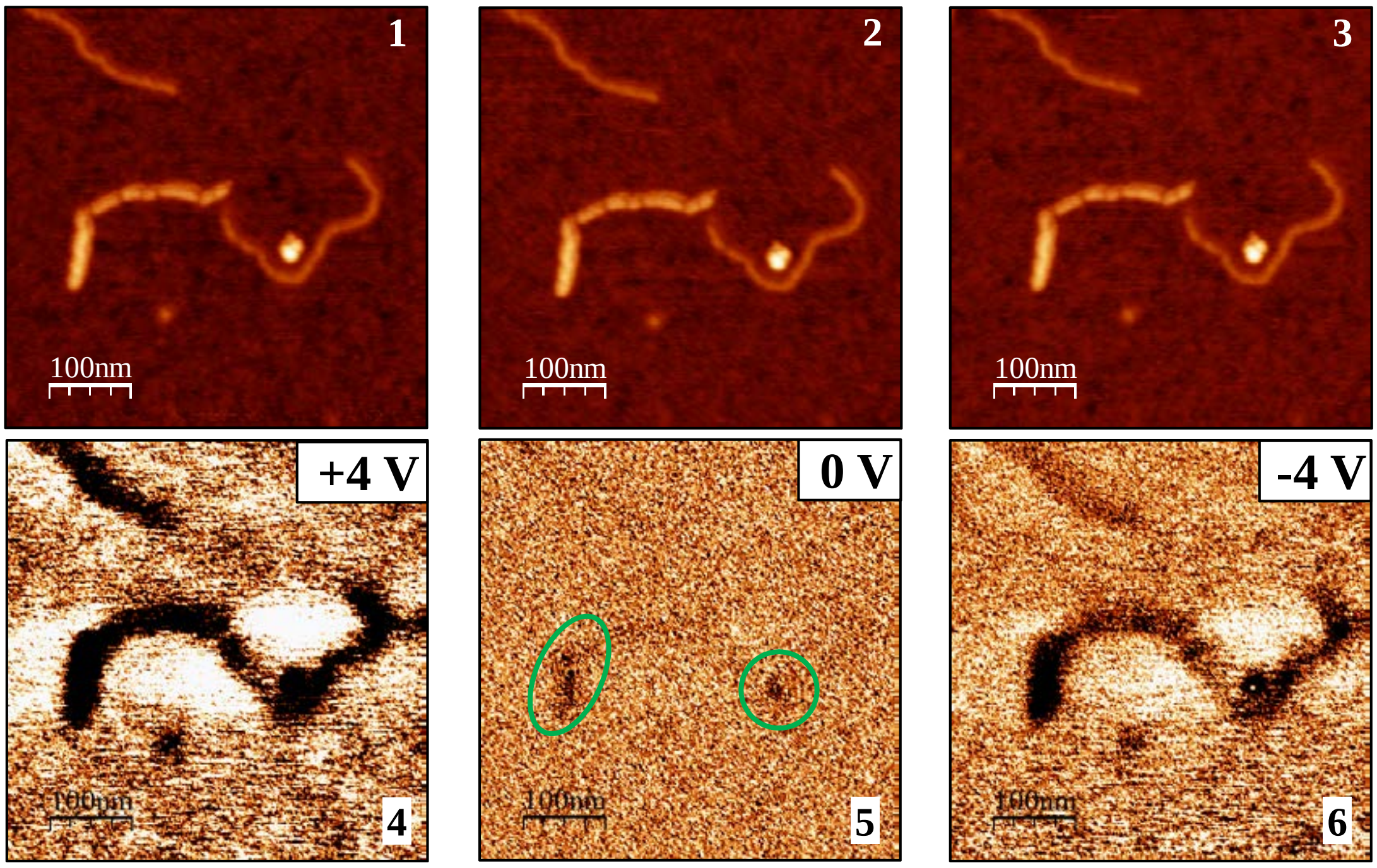

**Panel E: Ratio of intensities**

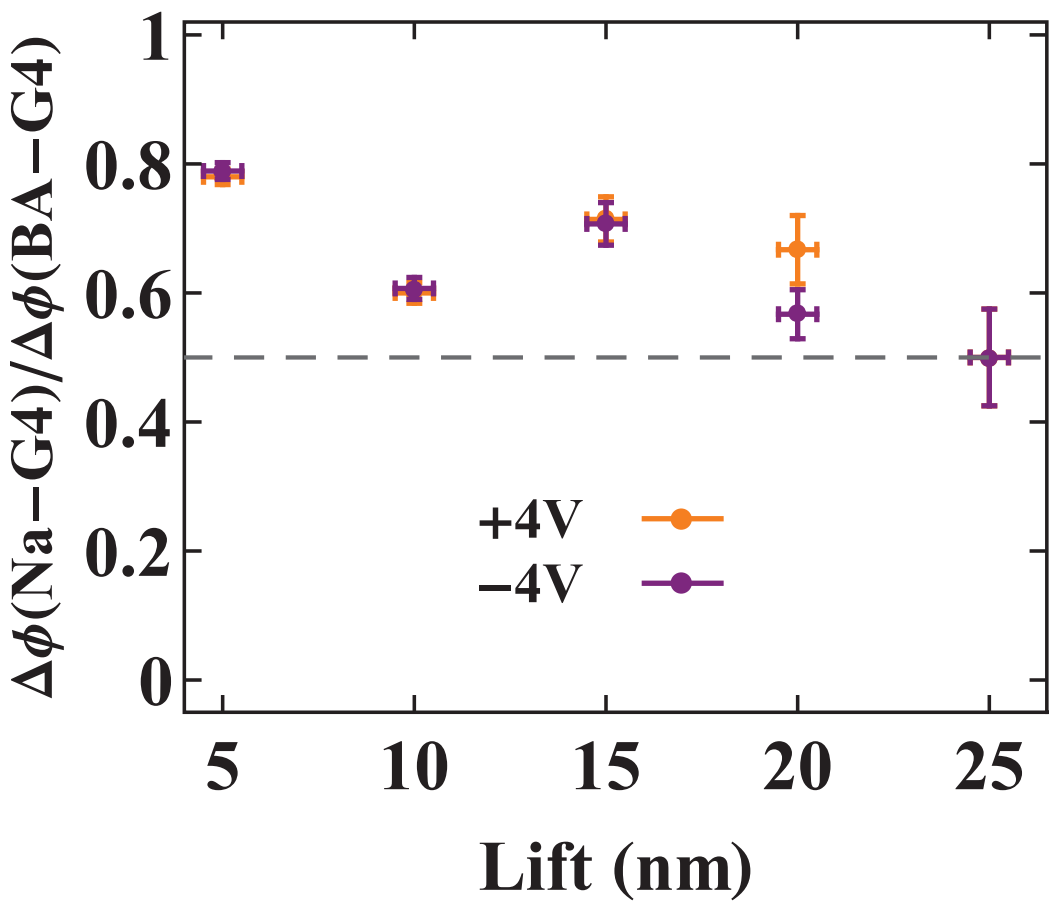

In this series of measurements, the amplitude was compensated following a slightly different approach. At the higher lift (20 nm), as in Figure S3, the amplitude during retrace was adjusted to be the same value as the amplitude at set-point height, ~10 nm. This enable greater sensitivity to the electrostatic forces at higher lifts. As the tip was lowered gradually, the possibility of VdW interactions became more substantial. To reduce this unwanted affect, the amplitude during retrace (at a bias of ±4 V) was adjusted to be just half of its value at set-point height, *i.e.* just ~5 nm. Deflection-distance and Amplitude-distance curves were measured at both set-point and lift height to ensure the height of the tip and the value of its amplitude. The results clearly indicate that regardless of the lift or size of amplitude tetra-molecular BA-G4-DNA possesses a stronger EFM signal compared to intra-molecular G4-DNA. A sample of the data at -4 V is shown.

**Figure S4.** EFM insensitivity to the amplitude. (a) - (d) a series of EFM measurements, taken in retrace mode. (a) AFM image of co-deposited long tetra-molecular BA-G4-DNA (top, left) and intra-molecular G4-DNA (bottom, right) taken in retrace mode at set-point height. A segment of a second intra-molecular G4-DNA molecule is visible on the right. Inset shows cross-sections of the molecules corresponding to the colored lines in the image, BA-G4-DNA (red) and G4-DNA (green). (b) EFM phase retrace at a nominal lift of 20 nm above set-point, measured at a bias of -4 V. A visible signal of BA-G4-DNA matches its location in (a). (c) Plot of the tip deflection as a function of the tip-substrate separation, *z*, (F-z), taken in retrace mode at the set-point height (blue) and at the nominal lift of 20 nm above set-point (green) at a bias of -4 V. (d) Plot of the tip amplitude as a function of *z*, revealing the sigmoidal shape of the curve at the lift, resulting from the application of bias and the electrostatic interaction (green), while at set-point height and at zero bias (blue) the curve is linear. The amplitude was adjusted during the lift to compensate for electrostatic attraction, and was set equal to the set-point amplitude. As a result only a slight deviation (18.5 nm vs. nominal 20 nm) was observed in the F-z curve. These measurements allow a determination of the absolute distance of the tip to the substrate. In this case, a 20 nm lift corresponds to 28 nm above the substrate. Similar series of EFM measurements, at a bias of -4 V, are shown in sets (e)-(h), (i)-(l) and (m)-(q), corresponding to nominal lifts of 15 nm, 10 nm and 5 nm, respectively. In these measurements, the amplitude was adjusted during retrace to equal nearly half of its set-point value, as is clearly seen in plots (h), (l) and (q). (r) Plot of the ratio of the averaged EFM signal of intra-molecular to tetra-molecular G4-DNA obtained for both negative and positive bias. This ratio is clearly less then unity. At the low lifts (5-10 nm) the value is ~0.35, while at the higher lifts (15-25) the value is ~0.5-0.6. A dashed line indicates a constant value of 0.5 for comparison. Error bars in position were determined from the F-z curves and system stability, while the error in the EFM signal is determined from the noise level of the signal.

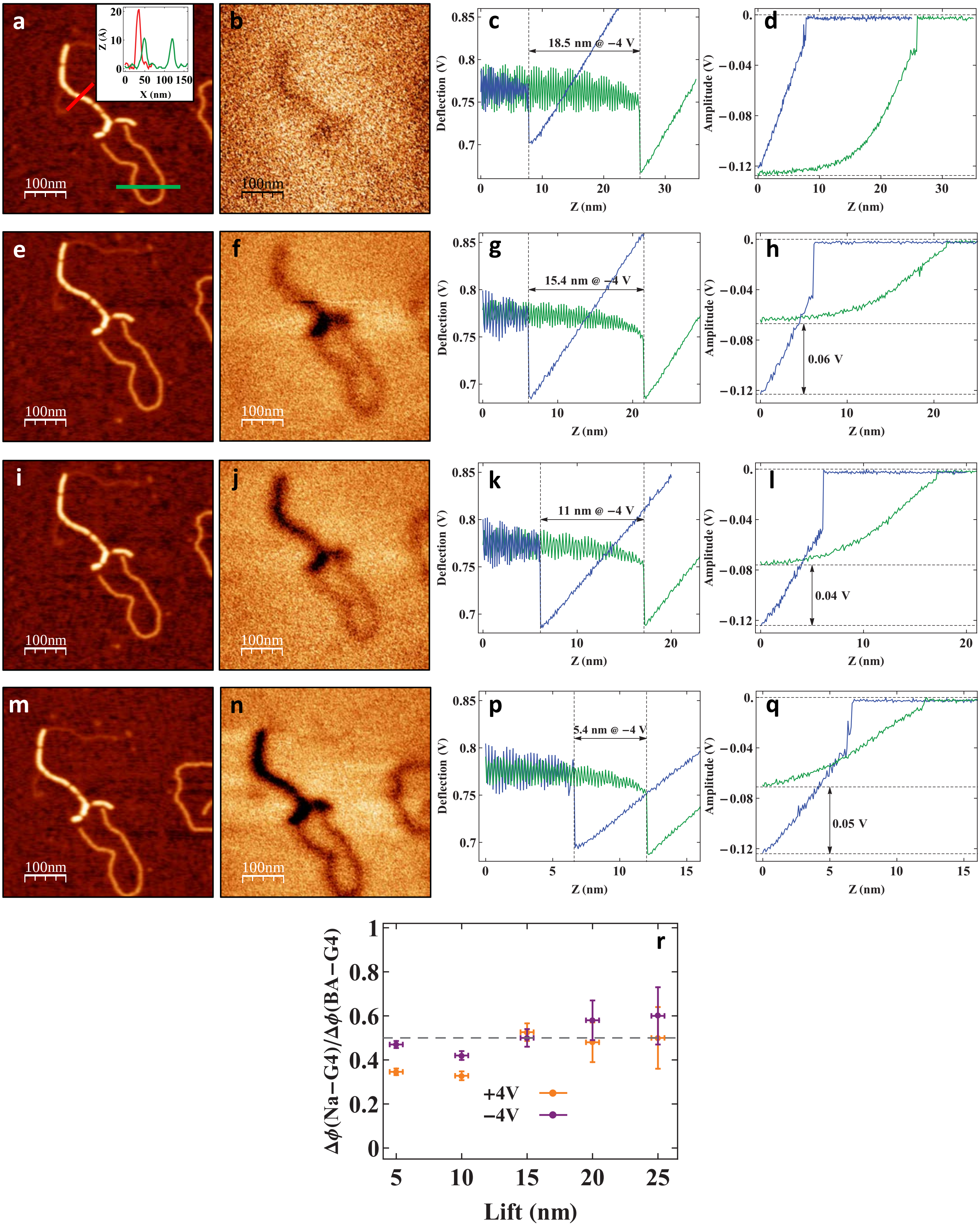

[1] J. T. Davis, Angew Chem Int Edit 2004, 43, 668.